# Should Classical Physics Be Interpreted Indeterministically?

Wyman Kwok[1]
[1]*Independent Researcher, PhD*

An indeterministic interpretation of classical physics has been proposed recently, in which the argument relies on attacking an alleged unwarranted metaphysical hidden assumption of the standard deterministic interpretation. This short paper aims at showing that it is arguably a strawman attack.

## I. INTRODUCTION

An argument for an indeterministic interpretation of classical physics (i.e., Newton's mechanics and Maxwell's electrodynamics) was put forth in Ref. [1] (see also [2] [3] [4] and [6] ). It maintained that although classical physics has traditionally been construed as deterministic (i.e., the physical laws determine a unique definite future (and past) state of a physical system once its current state is fixed, as famously revealed in the scenario of "Laplace's Demon"), it is not necessarily the case. There are metaphysical assumptions behind the traditional deterministic interpretation, and it is possible to give an alternative indeterministic interpretation by revising those assumptions, it contended. In particular, the usual practice that real numbers are used to represent physical quantities was held to be problematic, because this would lead to the unacceptable consequence of "infinite information density" (as related to the infinite string of digits following the decimal point of a real number) in the relevant physical space, according to it. In this paper, I argue that its construal of the standard deterministic interpretation is flawed, and therefore it arguably attacked a strawman.

## II. A STRAWMAN ATTACK?

According to Ref. [1], classical physics' "alleged deterministic character is based on the metaphysical, unwarranted assumption of 'infinite precision'" (p. 2), which can be formulated as a principle:

> *Principle of infinite precision*:
> (1) (Ontological)—there exists an actual value of every physical quantity, with its infinite determined digits (in any arbitrary numerical base).
> (2) (Epistemological)—despite it might not be possible to know all the digits of a physical quantity (through measurements), it is possible to know an arbitrarily large number of digits. (ibid.)

It claimed that "the limits of this principle rely on the faulty assumption of granting a physical significance to mathematical real numbers" and "[r]elaxing [this] assumption […] allows one to regard classical physics as a fundamentally indeterministic theory, contrarily to its standard formulation" (ibid.).

What did it mean by "granting a physical significance to mathematical real numbers"? It is just the content of the ontological part of the *principle of infinite precision*, meaning that some real numbers, with their infinite string of digits, are regarded as physically existing. And it contended that "[t]he physical existence of an infinite amount of predetermined digits could lead to unphysical situations, such as […] infinite information density" (p. 4). Those situations are unphysical because "an infinite amount of predetermined digits" has an infinite amount of information, but "finite volumes can contain only a finite amount of information" (p. 2).

Nevertheless, does the standard deterministic interpretation of classical physics really have the ontological part of the *principle of infinite precision* as its hidden assumption? First of all, let us ask, what is the "actual" value of the length of, say, certain pencil?[1] But obviously the value of its length depends on the unit (e.g. centimeter or inch) used to measure it, and there are in principle infinitely many "possible" units that can be used, and therefore there are only infinitely many possible values for its length but not a unique ontologically actual value. So, (1) is strange.

Regardless of why Ref. [1] maintained that the standard interpretation needs to make this unreasonable assumption, I contend that this ontological assumption should be replaced by the following one:

> (1') (Ontological)—there exists an actual definite physical state of the Universe at each moment, which can be characterized by physical quantities with representations in real numbers.

Note that a key here is that we make the distinction between "the physical quantities themselves" and "their representations in real numbers". For example, a pencil occupies space and has a physical quantity length. Suppose under certain measuring unit, this length is represented by a number x. According to Ref. [1]'s logic, here a commitment to the physical existence of x is in order. But this is just a confusion of the physical quantity itself and its number representation—what is physically real is the length itself but not its particular representation x. Afterall, with a different unit used, the length would be represented by a different number y, then does this mean that y is also physically real? Since there are in principle infinitely many possible units that can be chosen, does this mean that the infinitely many associated numbers are all physically real? Seemingly unreasonable. As a matter of fact, numbers are usually taken as abstract objects (which are neither in space nor in time), and a viewpoint regarding them as physically real probably results from conceptual confusion, unless the viewpoint is a sophisticated philosophical one which puts forth a controversial view of the ontological status of numbers, which then requires much justification. Ref. [1]'s case is not this because it talked about the physical existence of numbers quite unhesitatingly, seemingly without realizing the huge conceptual gap involved.

---

[1] Ref. [1] (p. 4) referred to Ref. [5] for a detailed explanation of a case in which a pencil's length involves infinite information density. My analysis below argues against this explanation.

How would the authors of Ref. [1] look at (1')? Obviously what they concerned most is the infinite amount of information associated with real numbers. Where is it in (1')? At least apparently, it is in the number representations but may not in the physical quantities themselves. So there may not have the sort of problem of infinite information density suggested by them. But can (1') do the job for determinism? At least not obviously not. An actual definite physical state of the Universe at one moment, together with the laws of nature, can determine other actual definite physical states at other moments. On the other hand, suppose under certain measuring unit, the length of a pencil is, say, $\sqrt{2}$. It does not mean that we require $\sqrt{2}$ to be physically existing, but only that one of the length's representations is $\sqrt{2}$. It also means when we measure the length in this unit, the actual value obtained would perfectly agree with $\sqrt{2}$, up to the accuracy of that measurement, say, 4 digits after the decimal point. And this measured value has only finite information. One might ask: if $\sqrt{2}$, as the length's representation, is not required to be physically real, how could the actual measurement agree with its value? But why not? Because by assumption, $\sqrt{2}$ is its length under this unit. If the measured value does not agree with this, then the assumption would be violated. A clearer question is in order so as to discuss the issue further. Also, be reminded that my foregoing analysis suggests that it is not reasonable to assume any of these number representations to be physically real (quick recall: 1. infinite possible values because of infinite possible choices of units; 2. the abstractness of numbers). Hence Ref. [1] would beg the question if it takes for granted that the standard deterministic interpretation must take this unreasonable move of reification of numbers. More arguments are needed.

### III. CONCLUDING REMARKS

Surely, there are many important remaining questions: What are physical quantities themselves? How are they related to their representations in numbers? What is the information content of a physical quantity itself? And how is this related to the information content of its number representations? Would infinite information density still be involved in (1') even though representations in real numbers are not taken as physically real? This short paper has not tackled these questions. What I have tried to show is: Ref. [1] arguably attacked a strawman when it assigned (1) to the standard deterministic interpretation of classical physics. What is needed to be shown is that (1') still has the problem of infinite information density (or other problems), with all the proper conceptual distinctions born in mind.

### ACKNOWLEDGEMENTS

I would like to thank Prof. Del Santo and an anonymous referee of Physical Review A for their useful comments on the drafts of this paper. Also, my gratitude goes to the participants of an on-line seminar in which I presented Ref. [1] and my responses to it, for the meaningful discussion. In particular, my friend Dr. Haipeng Zhang, who mentioned there the distinction between the thing itself and its representation, which might inspire me to introduce this distinction in this paper, although we referred to different things when using it.